\documentclass{aims}
  \usepackage{paralist}
  \usepackage{graphics} 
\usepackage{epstopdf}
 \usepackage[colorlinks=true]{hyperref}

 \usepackage{mathrsfs}
\usepackage{amssymb}
\usepackage{cite}
\usepackage{graphicx}
\usepackage{epsfig}
\usepackage{color}
\usepackage{amsfonts}
\usepackage{amsthm}
\usepackage{amsmath}
\newtheorem{remark}{Remark}
\newtheorem{prop}{\textbf{Proposition}}
\newtheorem{thm}{\textbf{Theorem}}
\newtheorem{cor}{\textbf{Corollary}}
\newtheorem{lem}{\textbf{Lemma}}
\def\eq#1{\begin{equation}#1\end{equation}}

\newcommand{\image}{{\rm Im\;}}

\def\n{{\bf n}}
 \def\qed{ \rule{.1in}{.1in}}
\hypersetup{urlcolor=blue, citecolor=red}

\def\currentvolume{X}
 \def\currentissue{X}
  \def\currentyear{2015}
   \def\currentmonth{XX}
    \def\ppages{X--XX}
     \def\DOI{10.3934/xx.xx.xx.xx}

\title[Decentralized gradient algorithm for linear equations] 
      {Decentralized gradient algorithm for solution of a linear equation}

\author[Anderson and Mou and Morse and Helmke]{}


 \email{brian.anderson@anu.edu.au}
 \email{mous@purdue.edu}
 \email{as.morse@yale.edu}
 \email{helmke@mathematik.uni-wuerzburg.de}

\thanks{This work of Brian D. O. Anderson is supported
by Australian Research Council Discovery Project DP-130103610, by NICTA (National ICT Australia), and by a DAAD-GO8 collaborative grant. The research of A. Stephen Morse is supported by the US Air Force Office of Scientific Research and by the National Science Foundation.The research of U. Helmke  has been supported by the grants HE 1858/13-1 (German Research Foundation) and 57139792 (DAAD-ARC Go8).}

\thanks{$^*$ Corresponding author: Brian D. O. Anderson}

\begin{document}
\maketitle

\centerline{\scshape Brian D. O. Anderson$^*$}
\medskip
{\footnotesize
 \centerline{College of Engineering and Computer Science }
   \centerline{The Australian National University}
   \centerline{ Canberra, Australia}
} 

\medskip

\centerline{\scshape Shaoshuai Mou}
\medskip
{\footnotesize
 \centerline{ School of Aeronautics and Astronautics}
   \centerline{Purdue University}
   \centerline{West Lafayette, IN, USA}
}

\medskip

\centerline{\scshape A. Stephen Morse}
\medskip
{\footnotesize
 \centerline{Department of Electrical Engineering}
   \centerline{Yale University}
   \centerline{New Haven, CT, USA}
}

\medskip

\centerline{\scshape Uwe Helmke}
\medskip
{\footnotesize
 \centerline{Department of Mathematics}
   \centerline{W{\"u}rzburg University}
   \centerline{W{\"u}rzburg, Germany}
}

\bigskip


\begin{abstract}
The paper develops a technique for solving a linear equation $Ax=b$ with a square and nonsingular matrix $A$, using a decentralized gradient algorithm. In the language of control theory, there are $n$ agents, each storing at time $t$ an $n$-vector, call it $x_i(t)$, and a graphical structure associating with each agent a vertex of a fixed, undirected and connected but otherwise arbitrary graph $\mathcal G$ with vertex set and edge set $\mathcal V$ and $\mathcal E$ respectively.  We provide differential equation update laws  for the $x_i$ with the property that each $x_i$ converges to the solution of the linear equation exponentially fast. The equation for $x_i$ includes additive terms weighting those $x_j$ for which vertices in $\mathcal G$ corresponding to the $i$-th and $j$-th agents are adjacent. The results are extended to the case where $A$ is not square but has full row rank, and bounds are given on the convergence rate.
\end{abstract}

\section{Introduction}

Among the contributions of John Moore was a significant body of work,
largely conducted with the last author of this paper, in which methods of control theory
were applied to provide algorithms solving various problems of linear
algebra, for example, matrix diagonalization. Typically, differential
equations were constructed whose solutions evolved in time to a steady
state containing the desired result. For example, for a given square
matrix, the equation might be initialized by the matrix and converge
to a steady state solution that is a diagonal matrix with eigenvalues
equal to those of the given matrix.  Many of these results are set
out in the book \cite{UJ94Book}.

This paper offers another contribution along these lines: we show how
to solve a linear equation $Ax=b$ in a \emph{distributed} way by a network of agents. Different from parallel algorithms in computer science  \cite{O96Cam,RJ05Cam,CAnd97Report} which usually require a common shared memory, a major novel feature of distributed algorithms lies in their implementation on multi-agent networks in which agents are provided with local memories and can communicate with each other. In the parallel world, the graph is chosen by the designer to maximize efficiency of computation in some sense whereas  in ours it may be given by other considerations such as communication requirements. When a linear equation of interest involves millions or more unknowns, it is likely to be impossible for a single memory to even store the whole linear equation. Such large scale linear equations can easily arise in electromagnetic field problems, in which the boundary integral technique is employed to recover the solution of a differential equation in space by an appropriate integration of the solution on the two-dimensional boundary surface \cite{Edelman93}. Another potential problem with the shared memory in parallel algorithms is the data safety. Security concerns often arise due to the fact that different agents are usually not necessarily in the same domain of trust. For example, when a customer turns to cluster computers for the help of computations which contain sensitive information such as business financial records, personally identifiable health information, etc, the customer may not be willing to share all such information with computers in the cluster \cite{CKJQ13TPD}. One natural way to avoid these problems is by developing such distributed algorithms for multi-agent networks. Since agents may also be physically separated from each other, each agent typically is  able to communicate only with certain other nearby agents. There are typically communication constraints on the information flow across multi-agent networks, which consequently preclude centralized processing and result in great interest of distributed algorithms.

One direction for solving linear equations in a distributed way is by reformulating them as a distributed optimization problem and then trying to employ existing algorithms in \cite{JAM12TAC,AA14TAC,DJJ14TAC,TAA14TAC} to solve them. Rather than go through the intermediate step of problem reformulation, the authors of \cite{SA13ECC,SJA13Allerton,SJA14TAC} have recently proposed a distributed algorithm for directly solving linear equations based on a so-called \emph{agreement principle}, which was implicitly used in \cite{AAP10TAC}. Here is the key idea: each agent limits the update of its state vector to satisfy its private equation, which is part of the linear equation $Ax=b$; at the same time a control is developed to drive all agents' states to reach a consensus vector, which means that this consensus vector must be the solution of $Ax=b$.  Algorithms obtained along this direction were first shown to work for non-singular matrix $A$ on fixed undirected networks in \cite{SA13ECC} and then were generalized to handle time-varying networks in \cite{SJA14TAC}.

In contract to the discrete-time algorithm proposed in \cite{SJA14TAC}, this paper proposes a continuous time algorithm and in so doing provides a different perspective on solving a linear equation in a distributed manner.
  The particular update rule for agent $i$'s state is
\begin{equation}
\dot x_i=f_{ii}x_i+\sum_{j\in\mathcal N_i}f_{ij}x_j,\quad i\in \n
\end{equation}
where with $\n=\{1,2,\ldots,n\}$, $\mathcal N_i$ denotes the set of labels of agent $i$'s neighbors' from which agent $i$ receives information, $x_i$ is the state vector associated with agent $i$, and the particular form of the coefficients $f_{ij}$ will be given subsequently. Now a feature of the algorithm is that it is, using the language of control theory, a form of {\it{consensus}} algorithm. This means that the different $x_i(t)$ converge as $t\rightarrow \infty$ to a common value, which will be the solution of the equation $Ax=b$. However what is probably the most distingishing feature of this work is its use of a  basic  result  from differential geometry,   which is perhaps not so well known  in control theory, to give a more or less immediate proof  of the paper's main result. In particular we use the fact that a gradient flow algorithm associated with a real analytic function on a real analytic Riemannian
manifold  necessarily converges to a fixed point. This result is
perhaps better known for the case of gradient flow algorithms in Euclidean space, see \cite{Lojasie84};
its more general form on Riemannian manifolds goes back to work of \cite{KTA00Math}.

The structure of the paper is as follows. In the next section, we motivate the form of the differential equations and state the main result.   The following section proves the main result, and Section 4 offers some comments on the convergence rate. Section 5 contains remarks relevant to future research.

%

%
%
%
%

\section{Establishing the consensus differential equations}

Our starting point is the algebraic linear equation
\begin{equation}
Ax=b
\end{equation}
in which $A\in\mathbb R^{n\times n}$ is non-singular and $b\in\mathbb R^n$. We shall rewrite the equation with $a_i^{\top}$ denoting the $i$-th row of $A$ as
\begin{eqnarray}
a_1^{\top}x&=&b_1\\
a_2^{\top}x&=&b_2\\
&\vdots&\\
a_n^{\top}x&=&b_n
\end{eqnarray}
 We shall later impose the requirement that $A$ is nonsingular, but for the moment the requirement is not in force.

The second rewriting of equations can be interpreted as stating that $x$ is a member of $n$ different manifolds (here affine subspaces), viz $a_i^{\top}x=b_i$.  Our approach to finding $x$ relies on finding $n$ vectors $x_i(t)$ each of which belongs for all time to a particular manifold, in that
\begin{equation}
a_i^{\top}x_i(t)=b_i, \quad i\in \n
\end{equation}
and so moving the $x_i$ that they become more and more like each other. If and when they take a common value,  call it $x$, it is obvious that it must satisfy the equation $Ax=b$ (whether or not $A$ is invertible).

The basis for adjusting the $x_i(t)$ involves setting up a cost function which penalizes any differences between them. To this end, consider also a connected graph $\mathcal G=(\mathcal V,\mathcal E)$ with $n$ vertices, and each is associated with an $n$-vector $x_i$.  Consider also the following cost function:
\begin{equation}
V(x_1,x_2,\ldots, x_n)=\frac{1}{2}\sum_{(i,j)\in\mathcal E}\|x_i-x_j\|^2
\end{equation}

Clearly $V$ achieves a global minimum of zero if and only if all $x_i$ have a common value. (Connectivity of the graph is essential to conclude the `only if' statement).

Now let $\mathcal M_i$ denote the manifold $a_i^{\top}x_i=b_i$, and
let $\mathcal M=\mathcal M_1\times \mathcal
  M_2\times\cdots\times \mathcal M_n$. Formally, one can regard $\mathcal M$
as the set of vectors $\mathcal X$ in  $\mathbb
  R^{n^2}$ obeying $n$ scalar constraints of the form
$$[0,0,\ldots,a_i^{\top},0,\ldots,0]\mathcal X=b_i.$$ (One should think of $\mathcal X$ as a vector obtained by stacking the $x_i$).
The function $V$ is obviously defined on $\mathbb R^{n^2}$, but it is
also defined on the affine subspace $\mathcal M \subset
  \mathbb R^{n^2}$. We observe that $V$ is clearly
  convex. This implies that the restriction $V|{\mathcal M}$ of $V$
  to  $\mathcal M$ is convex too because the $M_i$ are affine spaces, which shows a very useful property
  of $V|{\mathcal M}$. In the following proposition, we state two easily established properties linking the cost function to the linear equation.

\begin{prop} Assume that the equation $Ax=b$ is
  solvable. \begin{enumerate}
\item[(i)] The restricted cost function $V:{\mathcal M}\to \mathbb
  R$ possesses a global minimum $\mathcal X \in \mathcal M$. It is
  unique if and only if $Ax=b$ has a unique solution.
\item[(ii)] The local
  minima of $V|{\mathcal M}$ coincide with the global minima, and
  these in turn coincide with the set of critical points of
  $V|{\mathcal M}$.
\end{enumerate}
\end{prop}

The second part (ii) in the above Proposition is an immediate consequence of standard properties of smooth
convex functions on vector spaces $\mathcal M$.

If we were to regard the function $V$ as defined on
 $\mathbb R^{n\times n}$, and sought to compute a gradient flow from an arbitrary initial point in an attempt to reach the minimum, there would result
\begin{equation}\label{eq:ordinarygrad}
\dot x_i=-\sum_{j\in \mathcal N_i}(x_i-x_j),\quad i\in \n
\end{equation}
While this would result in a steady state in which all $x_i$ had the
same value, they would not be guaranteed to remain on the original
manifold. To remedy this defect, we obtain the gradient of $V$ on the
manifold $\mathcal M$. The way this is done in the
  optimization literature \cite{UJ94Book,David69}, given that the manifolds
$\mathcal M_i$ and therefore the manifold $\mathcal M$ are embedded in
a Euclidean space, is to simply project the gradient onto the tangent
space of the manifold. Because  $\mathcal M=\mathcal M_1\times \mathcal M_2\times\cdots\times\mathcal M_n$, it is not hard to check that this is equivalent to projecting $\dot x_i$ onto the manifold $\mathcal M_i$, Because the manifold $\mathcal M_i$ is an affine subspace, the tangent space is the set of vectors $z\in\mathbb R^n$ for which $a_i^{\top}z=0$.  Now given an arbitrary $y\in\mathbb R^n$, its projection onto the tangent space of the manifold $\mathcal M_i$ is simply $P_iy$, where $P_i$ denotes the orthogonal projection matrix to the kernel of $a_i^{\top}$ and for non-zero $a_i$ one has
\begin{equation}
P_i=I-\frac{a_ia_i^{\top}}{a_i^{\top}a_i}
\end{equation}
where $P_i$ denotes the projection matrix.
More precisely, this means that replacing (\ref{eq:ordinarygrad}), we have for the gradient flow on the manifold $\mathcal M$
\begin{equation}\label{eq:projgrad}
\dot x_i=-P_i\sum_{j\in \mathcal N_i}(x_i-x_j),\quad i\in \n.
\end{equation}
Of course we also require that the trajectory begins on $\mathcal M$, i.e. we require
\begin{equation}\label{eq:initialcond}
a_i^{\top}x_i(0)=b_i
\end{equation}
It is easy to check using the differential equation that $a_i^{\top}\dot x_i(t)=0$, which implies the trajectory stays on the manifold.

And now we can state the main result:
\begin{thm}\label{Thm_main}
Consider the linear equation $Ax=b$ with $A\in\mathbb R^{n\times n}$ and
$b\in \mathbb R^n$. Consider also a connected graph $\mathcal
G=(\mathcal V, \mathcal E)$ with $n$ vertices, and let $\mathcal N_i$
denote the neighbor set of vertex $i$. If $A$ is nonsingular, the
equation set (\ref{eq:projgrad}) with the initial condition (\ref{eq:initialcond}) has the property that $x_i(t)\rightarrow x$ for all $i\in \n$ as $t\rightarrow\infty$. Moreover, convergence is exponentially fast.  If $A \in\mathbb R^{m\times n}$ for some $m$ and has full row rank, convergence occurs to a solution of $Ax=b$.
\end{thm}

The proof of Theorem \ref{Thm_main} will be given later. One might well imagine that in the course of solving the differential equation set, round-off or other errors could move the $x_i(t)$ of the relevant manifold $\mathcal M_i$. This can be accommodated by adding a further term to the equations which restores the trajectory towards the manifold. The equations remain linear in character.

\begin{cor}\label{Cor_1}
Adopt the hypotheses of the theorem, save that (\ref{eq:projgrad}) is replaced by
\begin{equation}\label{eq:modprojgrad}
\dot x_i=-P_i\sum_{j\in \mathcal N_i}(x_i-x_j)-\frac{a_i}{a_i^{\top}a_i}(a_i^{\top}x_i-b_i),\ i\in \n
\end{equation}
and let the initial condition now be free. Then the conclusions of the theorem continue to hold.
\end{cor}

The proof of Corollary \ref{Cor_1} will be given in the next section. Further generalization of (\ref{eq:projgrad}) and (\ref{eq:modprojgrad}) can be achieved by introducing scalar positive gain constants (varying with $i$) in the equations, thus (\ref{eq:modprojgrad}) could be replaced, for arbitrary positive $\alpha$ and $\alpha_i$, by
\begin{equation}\label{eq:doublemodprojgrad}
\dot x_i=-\alpha P_i\sum_{j\in \mathcal N_i}(x_i-x_j)-\alpha_i\frac{a_i}{a_i^{\top}a_i}(a_i^{\top}x_i-b_i), \ i\in \n
\end{equation}

\section{Proof of Main Results}

We start this section with the proof of Theorem \ref{Thm_main}. By way of overview, we know from the general theory of
 real analytic gradient flows outlined in \cite{KTA00Math} that convergence must occur to an equilibrium point of the equations, and that because the manifolds $\mathcal M_i$ and thus the manifold $\mathcal M$ are closed, the equilibrium point must lie in $\mathcal M$ and thus the equilibrium values of the $x_i$ must lie in each $\mathcal M_i$. The equilibrium is necessarily a critical point of $V$. Now Proposition 1 (ii) indicates that the only critical points of $V$ are the global minima of $V|\mathcal M$. Under the hypothesis that $A$ is square and nonsingular, or that it is has full row rank, there is an $x$ such that $Ax=b$. An equilibrium point of the equations is given by $x_i =x$ for all $i$, which means that $\mathcal X\in\mathcal M$  assumes the form $\mathcal X=[x^{\top},x^{\top},\dots ,x^{\top}]^{\top}$). Such a value of $\mathcal X$ gives rise to $V=0$, i.e. corresponds to a global minimum. Conversely, at a global minimum, we can argue that  $\mathcal  X$ must take the form indicated, and then it follows that $Ax=b$. For suppose, in order to obtain a contradiction, that at a global minimum, there held $\mathcal X=[x_1^{\top},x_2^{\top}\ldots,x_n^{\top}]^{\top}$, and $x_i \neq x_j$ for some pair $ij$. Because of the connectedness of the graph $\mathcal G$, there is a path of edges in $\mathcal E$ connecting vertex $i$ to vertex $j$. Since $x_i \neq x_j$, there must hold for some edge, call it $rs$, along this path that $x_r\neq x_s$, and then immediately $V$ is seen to be nonzero since $\|x_r-x_s\|^2$ is one of the summands making up $V$.
Thus $V$ does not attain its global minimum. Hence a necessary and sufficient condition for any equilibrium point to which the equations converge is that consensus is attained, i.e. $x_i=x_j$ for all $i,j$ and the common value is a solution of $Ax=b$.

 We now give a more explicit algebraic derivation of the same conclusion, which will be useful in pinning down the exponential rate of convergence.
Let $L$ denote the Laplacian matrix associated with $\mathcal G$; note
that the connectivity property for $\mathcal G$ implies that $L$ is
nonnegative definite symmetric, with one eigenvalue at the origin, and
the associated eigenspace is the span of $ {\bf 1}$,
  where ${\bf{ 1}}$ denotes the vector with entries all 1. Let $e_i$
denote the unit vector in $\mathbb R^n$ with 1 in the $i$-th position. Denote the
matrix formed by placing the  $x_i \in \mathbb R^n$ next to one another as
 \begin{equation}
X=[x_1\;x_2\dots x_n] \in \mathbb R^{n\times n}
\end{equation}
(This should be distinguished from $\mathcal X$, which is obtained by stacking the $x_i$). At the equilibrium, there holds
\begin{equation}
-\big[I-\frac{a_ia_i^{\top}}{a_i^{\top}a_i}\big]XLe_i=0, i=1,2,\dots,n
\end{equation}
This implies that
\begin{equation}
XLe_i=\lambda_ia_i, i=1,2,\dots,n
\end{equation}
for some scalars $\lambda_i$ (which may be zero). In turn, with  $\Lambda=\operatorname{diag}[\lambda_i]$, there results
\begin{equation}
XL=[a_1\;a_2\dots a_n]\Lambda
\end{equation}
Since $\bf{1}$ is in the kernel of $L$, there results
\begin{equation}
[a_1\;a_2\dots a_n]\left[\begin{array}{c}\lambda_1\\\lambda_2\\\vdots\\\lambda_n\end{array}\right]=0
\end{equation}
and because the vectors $a_1,a_2,\ldots,a_n$ are independent under the theorem hypothesis (whether or not $A$ is square), we have that the $\lambda_i$ are all zero, that $XL=0$ and therefore the columns of $X$ are identical, i.e. consensus holds.

Exponential convergence follows from the fact that the equations are linear and time-invariant. If convergence occurs, it is necessarily exponential. This completes the proof of Theorem \ref{Thm_main}, and we shall return to the question of the rate of convergence subsequently.

\bigskip

To prove the corollary, we let $e_i=x_i-x^*$, where $x^*$ is a constant vector such that $Ax^*=b$. From $\dot{e}_i=\dot{x}_i$ and (\ref{eq:modprojgrad}) one has
 \begin{equation}\label{eq_erro0}
\dot e_i=-P_i\sum_{j\in \mathcal N_i}(x_i-x_j)-\frac{a_i}{a_i^{\top}a_i}(a_i^{\top}x_i-b_i),\ \ i\in \n
\end{equation} which together with $b_i=a_i^{\top}x^*$ implies $$\dot e_i=-P_i\sum_{j\in \mathcal N_i}\left((x_i-x^*)-(x_j-x^*)\right)-\frac{a_i}{a_i^{\top}a_i}(a_i^{\top}x_i-a_i^{\top}x^*)$$ Thus \eq{\label{eq_error1}\dot e_i=-P_i\sum_{j\in \mathcal N_i}(e_i-e_j)-(I-P_i)e_i, \quad i\in \n}
To write the equations (\ref{eq_error1}) in a more compact form, we let $e=\left[
                                   \begin{array}{cccc}
                                     e_1' & e_2' & \cdots & e_n' \\
                                   \end{array}
                                 \right]'
$, $P=\operatorname{diag}[P_1,P_2,\dots, P_n]$ and $\bar L=L\otimes I_n$. One has \eq{\label{eq_error2}\dot e=-(P\bar{L}+I-P)e} Proving that all $x_i$ converge to $x^*$ exponentially fast is equivalent to proving that $e$ converges to 0 exponentially fast, for which we need the following lemma:
\begin{lem}\label{Lem_eig} If $\ker A=0$, all eigenvalues of $P\bar{L}+I-P$ are real and positive.
\end{lem}

To prove Lemma \ref{Lem_eig} we need the following lemma about eigenvalues.
\begin{lem}\label{Lem_eigP}
If $M$ and $N$ are square matrices of the same size, then $MN$ and $NM$ have the same eigenvalues.
\end{lem}
\noindent{\bf Proof of Lemma \ref{Lem_eigP}:} By Sylvester's Determinant Theorem \cite{Syl51}, one has \eq{\det(I-\frac{1}{\lambda}MN)=\det( I-\frac{1}{\lambda}NM)} for $\lambda\neq 0$. Thus $MN$ and $NM$ share the same non-zero eigenvalues. Moreover, by $\det(MN)=\det(M)\det(N)=\det(NM)$, one has if $0$ is an eigenvalue of $MN$ or $NM$, it must be also the eigenvalue of the other. Thus $MN$ and $NM$ share the same eigenvalues if they are both square of the same size. $\qed$

\noindent{\bf Proof of Lemma \ref{Lem_eig}:} Let us first establish that the eigenvalues of $P\bar L+I-P$ are identical with those of $P\bar LP+I-P.$ (We will then analyse the eigenvalues of the latter matrix) To see this, observe that because $P^2=P$, $$P\bar L-P=P^2\bar L-P^2  =P(P\bar L-P)$$
Hence the eigenvalues of $P\bar L-P$ are the same as those of $(P\bar L-P)P=P\bar L P-P$ by Lemma \ref{Lem_eigP}. Consequently, the eigenvalues of $P\bar L+I-P$ are the same as those of $P\bar LP+I-P$. Observe first that $P\bar LP+I-P$ is positive semi-definite since $P\bar L P$ and $I-P$ are positive semi-definite. Thus to prove Lemma \ref{Lem_eig}, it is sufficient to establish that $P\bar LP+I-P$ is non-singular. We will prove this in the following by assuming the contrary to obtain a contradiction.

Suppose there is a nonzero $\alpha$ for which $[P\bar LP+(I-P)]\alpha=0$. Then clearly
\begin{eqnarray}
\bar LP\alpha&=&0\\
(I-P)\alpha&=&0
\end{eqnarray}
It follows that $\bar L\alpha=0$ and so $\alpha={\bf{1}}\otimes q$ for some $q\in \mathbb R^n$. Then again from the fact that $(I-P)\alpha=0$, we obtain $a_i^{\top}q=0$ for all $i\in \n$, whence $q$ and thus $\alpha$ is zero. Then $P\bar{L}P+I-P$ is non-singular. $\qed$
\bigskip

The fact that exponentially fast convergence occurs means that if $b$ is varying, the algorithm will still lead to an approximate consensus solution of $Ax=b$, with the quality of the approximation linked to the rate of variation of $b$.

\section{Convergence rates}
There is an alternative way of writing the equations (\ref{eq:projgrad}) in the following compact form
\begin{equation}\label{eq_x0}
\dot{\mathcal X}=-P\bar L\mathcal X
\end{equation} where $\mathcal{X}=\left[
                                   \begin{array}{cccc}
                                     x_1^{\top} & x_2^{\top} & \cdots & x_n^{\top} \\
                                   \end{array}
                                 \right]^{\top}
$.
Evidently, when the $a_i$ are linearly independent, there are $n$ eigenvalues of $P \bar L$ at the origin, and the remainder have negative real parts. The rate of exponential convergence of the system (\ref{eq_x0}) to its equilibrium is determined by the smallest non-zero eigenvalue of $P\bar{L}$ which we shall denote by $\rho$.

Note that $P\bar{L}$ and $P\bar{L}P$ have the same eigenvalues because $P\bar{L}=P(P\bar{L})$ and $P(P\bar L)$ and $(P\bar L)P$ have the same eigenvalues. Thus $\rho$ is real and positive and is equal to the smallest non-zero eigenvalue of $P\bar{L}P$. In order to find a bound for $\rho$, we will study non-zero eigenvalues of $P\bar{L}P$.

%
%

Let $Q=\left[
                                   \begin{array}{cc}
                                     \hat{Q} & \bar{Q} \\
                                   \end{array}
                                 \right]$ be a square orthogonal matrix such that the columns $\hat{Q}$ and $ \bar{Q}$ form a basis for $\ker P$ and $\image P$, respectively.
Then \begin{eqnarray}Q^{\top}P\bar{L}PQ&=&\left[
                        \begin{array}{c}
                          0 \\
                          \bar{Q}^{\top}P \\
                        \end{array}
                      \right]\bar{L}\left[
                                      \begin{array}{cc}
                                        0 & P\bar{Q} \\
                                      \end{array}
                                    \right]\\
                                    &=&
\left[
                        \begin{array}{cc}
                          0 & 0 \\
                          0 & \bar{Q}^{\top}P\bar{L}P\bar{Q}\\
                        \end{array}
                      \right]\\
                      &=& \left[
                        \begin{array}{cc}
                          0 & 0 \\
                          0 & \bar{Q}^{\top}\bar{L}\bar{Q}\\
                        \end{array}
                      \right]\label{eq_QQ}
\end{eqnarray} The last equality (\ref{eq_QQ}) comes from $P\bar{Q}=\bar{Q}$ since the columns of $\bar{Q}$ forms a basis for $\image P$ and $P^2=P$.

Thus all non-zero eigenvalues of $Q^{\top}P\bar{L}PQ$ are the same as those of the non-singular matrix $\bar{Q}^{\top}\bar{L}\bar{Q}$. From $\bar{Q}^{\top}\bar{Q}=I_{n(n-1)}$ and the Poincare Separation Theorem \cite{RC85Book}, one has \eq{\label{eq_rho1}\lambda_1(\bar{L})\leq \lambda_1(\bar Q^{\top} \bar L \bar Q)\leq \lambda_{n+1}(\bar L)} where $\lambda_j(\cdot)$ denote the $j$th smallest eigenvalue of a Hermitian matrix. Since $\bar{L}=L\otimes I_n$ and $L$ is the Laplacian of a connected graph, one has \eq{\label{eq_rho2}\lambda_1(\bar L)=0,\quad \lambda_{n+1}(\bar L)=\lambda_2(L)}
Recall that $\rho$ is equal to the smallest non-zero eigenvalue of $P\bar{L}P$, which is similar to $Q^{\top}P\bar{L}PQ$. Then \eq{\label{eq_rho3}\rho=\lambda_1(\bar{Q}^{\top}\bar{L}\bar{Q})} From (\ref{eq_rho1}) to (\ref{eq_rho3}), one reaches a trivial lower bound for $\rho$, which is 0, and the following upper bound:
\begin{thm}
The smallest non-zero eigenvalue of $P\bar{L}$ is upper bounded by \eq{ \rho \leq \lambda_2(L)}
\end{thm}
\begin{remark}
$\lambda_2(L)$ is called the algebraic connectivity of a graph. It is bounded below by \cite{Mohar91} \eq{\label{eq_upL}\lambda_2(L)\geq \frac{4}{nD}} with $D$ the diameter of the graph, and is bounded above by  $$\lambda_2(L)\leq \frac{n}{n-1}$$ for $n\geq 2$ with equality holding if and only if the graph is complete \cite{Fan97}. This further gives an upper bound of $\rho$ in terms of the number of agents in the network.

From (\ref{eq_upL}) one observes that $\lambda_2(L)$ could be very small, which suggests $\rho$ may be close to 0 for certain graphs. Moreover, (\ref{eq_rho3}) implies that $\rho$ is related to both $\bar{L}$ and $\bar{Q}$, the latter of which is determined by the matrix $A$. It may be impossible to obtain a non-trivial lower bound for $\rho$ without assuming more about $A$ beyond non-singularity.
\end{remark}
\section{Conclusions}

There are a number of issues that are related to the ideas of this paper, but remain unexplored. We comment briefly on some of them.

In case the matrix $A$ is tall, in general the equation $Ax=b$ cannot be solved, but it does make sense to search for a least squares solution. Theoretically, this could be obtained in the full rank case by working with the linear equation $A^{\top}Ax=A^{\top}b$, but any such approach requiring the initial computation of $A^{\top}A$ and $A^{\top}b$ might be contrary to the spirit of seeking a decentralized solution because of  the associated computations.  For example, in distributed parameter estimation \cite{SJK12IT}, a multi-agent network aims to solve a group of observation equations $A_ix=b_i$. Because the $b_i$ are usually contaminated with measurement noise, the whole linear equation $Ax=b$ is usually overdetermined.  How then to obtain a least squares solution is for the moment an open problem. An alternative idea to obtain the least square solution in a distributed way was briefly mentioned in \cite{SJA14TAC} by solving a larger linear equation than $Ax=b$. However, this idea requires each agent to control an augmented state vector the dimension of which does not scale well with the number of agents in the network. This scaling problem has recently been partially ameliorated in \cite{SAZLD15SCL} by assuming a sparse structure for the linear equations of interest.

It would clearly be relevant to contemplate algorithms in which rows of $A$ were grouped together. The manifolds $\mathcal M_i$ would be defined by equations like $A_ix_i=b_i$ where $A_i$ was a fat matrix. The changes to the algorithm and the associated proof are trivial to contemplate. One issue is what the difference in computational burden might be.

Evidently, the fact that the only requirement on the graph is that it be connected allows the overlaying of a concept like sparseness for each of the equations for $x_i$, that is the differential equation for $x_i$ need only have at most one other $x_j$ feeding into it.  How to exploit sparseness in the matrix $A$ is less clear. Obviously though, inner products involving $a_i$ are easier to compute when $A$ is sparse.

In the introduction, we noted that a discrete form of the algorithm is available, see \cite{SA13ECC,SJA13Allerton,SJA14TAC}. It should be immediately derivable from the equations presented here. Whether there could be issues of stiffness that would cause problems in discretizing the equations is unknown.

In numerical linear algebra, the difficulty of executing certain calculations is frequently evaluated, and often the formula involves the dimension of the underlying matrices. It is evident here there are tradeoffs possible between convergence rate, storage requirements and complexity requirements in an implementation of a differential equation solver. How all these interact and what sort of comparison can be made with conventional counts for matrix inversion is unknown.

There are several other more significant directions in which this work might be developed. First, one could seek to add linear inequalities or more generally convex inequalities into the problem statement. Reference \cite{AAP10TAC}  can be thought of as  explaining the use of consensus for distributed optimization in discrete time, and its existence is prima facie evidence of the reasonableness of seeking to generalize the ideas of this paper. Second,  in coding theory there are sometimes requirements to solve linear equations with tall matrices whose entries are in a finite field. One might speculate  as to whether a consensus approach could work, and be of benefit, for such problems. Of course, the notion of projection onto manifolds would not be expected to play any role.
Last, we comment that the device of representing the equation to be solved as a consensus problem, and using a series of manifolds as a lens through which to view the problem, is a device that can find application to many control tasks. For example, consider the task of aligning three coordinate frames each in $\mathbb R^3$. This is a consensus problem on a sphere, and gradient flow on a sphere is easy to compute. Thus one can readily devise an algorithm to secure the alignment.

\medskip
Received xxxx 20xx; revised xxxx 20xx.
\medskip

\end{document}